\documentclass[useAMS,usenatbib]{mnras}
\usepackage[dvips]{graphicx}
\usepackage{epsfig}
\usepackage{rotating}
\usepackage{natbib}
\usepackage{color}
\usepackage{mathrsfs}
\usepackage{amssymb}
\usepackage[caption=false]{subfig}

\newcommand{\Msun}{\, {\rm M_{\odot}}}

\title[Morphology and the galaxy stellar-to-halo mass relation]{The dependence of the galaxy stellar-to-halo mass relation on galaxy morphology}
\author[C.A.~Correa \& J.~Schaye]
{Camila A.~Correa$^{1,2}$\thanks{E-mail: camila.correa@uva.nl} \& Joop Schaye$^1$\\
$^1$ Leiden Observatory, Leiden University, P.O. Box 9513, 2300 RA Leiden, The Netherlands\\
$^2$ Institute for Theoretical Physics Amsterdam, University of Amsterdam, Science Park 904, 1098 XH Amsterdam, The Netherlands}
\date{\today}

\pagerange{\pageref{firstpage}--\pageref{lastpage}} \pubyear{2019}

\def\LaTeX{L\kern-.36em\raise.3ex\hbox{a}\kern-.15em
    T\kern-.1667em\lower.7ex\hbox{E}\kern-.125emX}

\begin{document}
\maketitle

\begin{abstract}
We investigate the dependence of the local galaxy stellar-to-halo mass relation (SHMR) on galaxy morphology. We use data from the Sloan Digital Sky Survey DR7 with morphological classifications from Galaxy Zoo, and compare with the EAGLE cosmological simulation. At fixed halo mass in the mass range $10^{11.7}-10^{12.9}\Msun$, the median stellar masses of SDSS disc galaxies are up to a factor of 1.4 higher than the median masses of their elliptical counterparts. However, when we switch from the stellar masses from Kauffmann et al. to those calculated by Chang et al. or Brinchmann et al., the median SHMR from discs and ellipticals coincide in this mass range. For halo masses larger than $10^{13}\Msun$, discs are less massive than ellipticals in same-mass haloes, regardless of whose stellar mass estimates we use. However, we find that for these high halo masses the results for discs may be affected by central/satellite misclassifications. The EAGLE simulation predicts that discs are up to a factor of 1.5 more massive than elliptical galaxies residing in same-mass haloes less massive than $10^{13}\Msun$, in agreement with the Kauffmann et al. data. Haloes with masses between $10^{11.5}$ and $10^{12}\Msun$, that host disc galaxies, were assembled earlier than those hosting ellipticals. This suggests that the discs are more massive because they had more time for gas accretion and star formation. In $10^{12}-10^{12.5}\Msun$ haloes, the central black holes in elliptical galaxies grew faster and became more massive than their counterparts in disc galaxies. This suggests that in this halo mass range the ellipticals are less massive because AGN feedback ejected more of the halo's gas reservoir, reducing star formation, and suppressing the (re)growth of stellar discs.
\end{abstract}


\section{Introduction}

A central ansatz in the $\Lambda$CDM cosmological paradigm is that galaxies form from baryonic condensations within the potential well of a dark matter halo (e.g., \citealt{White78}). The baryonic physics that leads to the formation of galaxies is complex, it involves gravitational instabilities, gas heating, cooling and dissipation, galaxy-galaxy mergers and interactions, feedback from supernovae and black holes. Therefore, the physical and statistical connection between galaxies and dark matter haloes, commonly called the galaxy-halo connection (see e.g. \citealt{Wechsler18} for a recent review), is essential to our understanding of the galaxy formation process in a cosmological context. 

The typical galaxy stellar mass at a given halo mass, or galaxy stellar-to-halo mass relation, which we hereafter abbreviate as SHMR, has been extensively studied using various observational techniques.  Galaxy-galaxy lensing uses distortions of the shapes and orientations of background galaxies caused by intervening mass along the line of sight to infer the foreground mass distribution in stacks (e.g. \citealt{Zu15,Zu16,Mandelbaum16,Leauthaud17,Sonnenfeld18}). Satellite kinematics uses satellite galaxies as test particles to trace out the dark matter velocity field, and thus the potential well, of the dark matter halo (see e.g., \citealt{More11,Wojtak13,Lange19,Tinker19}). Other approaches, such as abundance matching (e.g., \citealt{Guo10,Behroozi13,Moster13}) and galaxy clustering (e.g., \citealt{vandenBosch07,Zheng07,Hearin13,Guo16,Zentner16}), compare the observed abundance and clustering properties of galaxy samples with predictions from a phenomenological halo model. 

Constraints on the SHMR from these different methods (e.g. \citealt{Yang09,Guo10,Wang10,Reddick13,Behroozi13,Moster13,Moster18,Kravtsov18,Behroozi19}) have shown that the stellar mass ($M_{*}$) of central galaxies scales as $M_{*}\propto M_{h}^{2-3}$ at dwarf masses (with $M_{h}$ the halo mass) and as $M_{*}\propto M_{h}^{1/3}$ at cluster masses. However, the dependence of the SHMR for central galaxies on the galaxies' properties, such as morphology and color, is not yet fully understood. 

Galaxies in the local Universe tend to be either blue star-forming discs or red passive ellipsoids, and can thus be divided into two distinct populations based on their optical color and morphology (e.g., \citealt{Strateva01,Baldry04,Willett13}). \citet{Mandelbaum16} investigated whether central passive and star-forming galaxies, which have different star formation histories, also have different relationships between stellar and halo mass. From a sample of locally brightest galaxies from the Sloan Digital Sky Survey (hereafter SDSS), and galaxy-galaxy lensing halo mass estimates, they reported that over the stellar mass range $10^{10.3}-10^{11.6}\Msun$ (halo mass range $10^{11.5}-10^{14}\Msun$) passive central galaxies have haloes that are at least twice as massive as those of star-forming objects of the same stellar mass. Although this was an exciting result, they observed large disagreement with other studies that used different analysis techniques such as a combination of satellite kinematics, weak lensing and abundance matching (\citealt{Dutton10}), satellite kinematics (\citealt{More11}), clustering and abundance matching (\citealt{Rodriguez15}), or empirical abundance modelling (\citealt{Hearin14,Moster19}), over a similar stellar and halo mass range. \citet{Mandelbaum16} concluded that large statistical or systematic uncertainties can make it difficult to draw a definitive conclusion. A similar conclusion was reached in the recent review of \citet{Wechsler18}. 

Despite this lack of consensus, \citet{Cowley19} attempted to constrain the SHMR of passive and star-forming galaxies at higher redshifts, in the range $z\approx 2-3$, as identified in the Spitzer Matching Survey of the UltraVISTA ultra-deep Stripes. They adopted a halo modelling approach and, opposite from \citet{Mandelbaum16}, they showed that at fixed halo mass, passive central galaxies tend to have larger stellar masses than their star-forming counterparts. They proposed that passive galaxies reside in haloes with the highest formation redshifts at a given halo mass.

Recently, \citet{Taylor20} use KiDS weak lensing data (\citealt{Hildebrandt17}) to measure variations in mean halo mass as a function of various galaxy properties, such as colour, specific star formation rate, Sersic index and effective radius, for a volume-limited sample of GAMA galaxies (\citealt{Driver11}). They concluded that for the stellar mass range $2-5\times 10^{10}\Msun$, size and Sersic index are better predictors of halo mass than colour or specific star formation, suggesting that the mean halo mass is more strongly correlated with galaxy structure than either stellar populations or star formation rate.

A complementary way to investigate the dependence of the SHMR on galaxy properties is to resort to cosmological simulations of galaxy formation. The current state of the art of such efforts comprises an N-body computation of the evolution of dark matter combined with either a hydrodynamical (e.g., \citealt{Vogelsberger14,Schaye15,Dubois16,Hopkins18,Nelson19,Dave19}), semi-analytical (e.g., \citealt{Croton16,Lacey16,Xie17,Cora18,Lagos18}) or parameterised (e.g., empirical modelling, \citealt{Mo96,Conroy06,Moster19,Behroozi19}) treatment of the baryonic processes involved. Although these theoretical approaches have been very successful at reproducing multiple observational data sets, they are still limited by our lack of knowledge regarding complex physical processes, such as stellar and black hole feedback processes (see e.g. \citealt{Davies20}), that directly impact on the galaxies' stellar mass. 

In a recent effort, \citet{Moster19} analysed the SHMR that resulted from the empirical model EMERGE, which was constrained by requiring a number of statistical observations to be reproduced. \citet{Moster19} showed that over the stellar mass range $10^{10.5}-10^{11.5}\Msun$ (halo mass range $10^{12}-10^{13.5}\Msun$), at fixed halo mass present-day early-type (or passive) galaxies are more massive than late-type (or star-forming) galaxies, whereas at fixed stellar mass early-type galaxies populate more massive halos, in agreement with lensing results. They concluded that this dependence arises from the scatter in the SHMR.

In this work we investigate how galaxy morphology and color affect the galaxy-halo connection, specifically the SHMR. We resort to the EAGLE simulation (\citealt{Schaye15,Crain15}) for this study, but also analyse a large SDSS DR7 (seventh data release) galaxy dataset, combined with the Galaxy Zoo DR1 data (\citealt{Lintott08,Lintott11}) to split galaxies by morphology, and with a group catalogue (\citealt{Yang07}) to split galaxies by halo mass and into centrals and satellites. 

This paper is organised as follows. In Section 2 we introduce the SDSS catalogue constructed for this study, analyse the completeness of the sample, and estimate the SHMR. We discuss the differences in the techniques used to measure galaxy stellar masses, as well as possible biases that may erase or be responsible for the morphology dependence of the SHMR in Section 2.4. Section 3 describes the EAGLE simulation and shows the SHMR dependence on morphology for EAGLE galaxies. Section 4 investigates the physical origin of the EAGLE morphology-SHMR. Finally, Section 5 summarises the main findings.

\section{SDSS Observations}\label{SDSS_section}

\subsection{Data}\label{SDSS_sample}

To investigate the impact of galaxies' color and morphology on the SHMR of local galaxies, we use the Sloan Digital Sky Survey (\citealt{York00}), Data Release 7 (\citealt{Abazajian09}), an extensive five passband ($u$, $g$, $r$, $i$ and $z$) imaging and spectroscopic survey. We cross-match the SDSS sample with the New York University Value-Added Galaxy Catalogue (NYU VAGC; \citealt{Blanton05,Padmanabhan08}), with the Max Planck Institute for Astrophysics John Hopkins University (MPA JHU; \citealt{Kauffmann03,Brinchmann04}) catalogue, as well as with the stellar mass catalogue from \citet{Chang15}. 

Stellar masses are calculated by multiplying the dust-corrected luminosities of galaxies with mass-to-light (M/L) ratios, with the latter being constrained from broadband photometry and spectral-fitting techniques. The relation between M/L ratio and galaxy color depends on metallicity, dust and star formation history (hereafter SFH), which can be determined by modeling broadband spectral energy distributions (SEDs) with stellar population synthesis. \citet{Brinchmann04} assumed exponentially decaying SFHs and performed fits to the SDSS photometry using \citet{Bruzual03} stellar population synthesis models. \citet{Chang15} combined SDSS and WISE photometry for the full SDSS spectroscopic galaxy sample, further adding mid-infrared emission tracers of star formation activity, and fitted the photometric SED using the software MAGPHY (\citealt{DaCunha08}) as well as \citet{Bruzual03} templates. 

\citet{Bell03} noted that some of the largest uncertainties in derived M/L ratios come from uncertainties in the assumed SFHs, in particular the presence of bursty star-forming episodes. \citet{Kauffmann03} used two stellar absorption-line indices, the 4000 $\AA$ break ($D_{n}(4000)$) and the Balmer absorption line index H$\delta_{A}$, to better constrain the SFHs and M/L ratios. The location of a galaxy in the $D_{n}(4000)-$H$\delta_{A}$ plane is a powerful diagnostic of whether the galaxy has been forming stars continuously or in bursts over the past 1-2 Gyr. They assigned stellar M/L ratios to their galaxies using a bayesian analysis to associate the observed $D_{n}(4000)$ and H$\delta_{A}$ values with a model drawn from a large library of Monte Carlo realizations of different SFHs. A comparison with broadband photometry yielded estimates of the dust attenuation. These stellar masses were calculated assuming a Kroupa (\citealt{Kroupa01}) initial mass function (IMF), we convert them to a \citet{Chabrier03} IMF by multiplying by a factor of 0.88 (\citealt{Cimatti08}). 

Throughout this work the stellar masses from \citet{Kauffmann03} are used unless stated otherwise. We assume a $\Lambda$CDM flat cosmology with $h=0.6777$ and $\Omega_{\rm{m}}=0.307$ (as derived by \citealt{Planck}), and multiply by $h^{2}$ or $h$ when necessary to remove the $h$ dependence.

\begin{figure} 
\includegraphics[angle=0,width=0.46\textwidth]{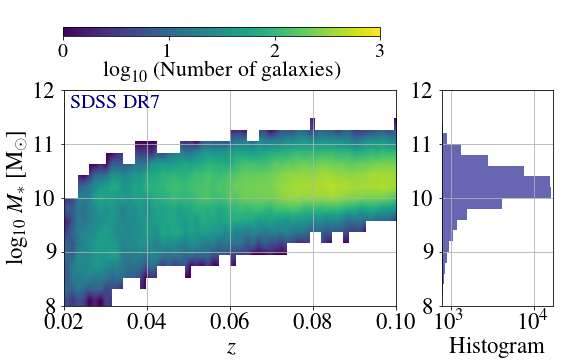}
\caption{Stellar mass as a function redshift for 127,780 SDSS galaxies that result from the cross-match of the group and morphology catalogues of Yang et al. (2007) and Lintott et al. (2011), respectively. The color scale indicates the number count of galaxies in a particular stellar mass and redshift bin (with lighter colors corresponding a higher number of galaxies). The distribution of the galaxy sample in stellar mass bins is shown in the right panel.}
\label{SDSS_sample}
\end{figure}

We cross-match the SDSS data with the galaxy group catalogue from \citet{Yang07} to extract halo masses and the central/satellite galaxy classifications. The galaxy group catalogue comprises galaxies in the range $0.02<z<0.20$ with a redshift completeness larger than 0.7. \citet{Yang07} did not measure halo masses directly, but rather {\it{estimated}} the masses by employing a halo-based group finder to iteratively determine the group membership of a galaxy based on a luminosity-scaled radius. In the first iteration, the adaptive halo-based group finder applies a constant mass-to-light ratio of 500 $h\Msun/L_{\odot}$ to estimate a tentative halo mass for each group. This mass is then used to evaluate the size and velocity dispersion of the halo embedding the group, which in turn are utilized to define group membership in redshift space. At this point, a new iteration begins, whereby the group characteristic luminosity and stellar mass are converted into halo mass using the halo occupation model of \citet{Yang05}. This procedure is repeated until no more changes occur in the group membership. In each group sample, galaxies are classified as centrals (the most massive group members in terms of stellar mass), and satellites (all other group members less massive than their group central).  

Dark matter halo masses, $M_{h}$, associated with the host groups were estimated on the basis of the ranking of both the group total characteristic luminosity and the group total characteristic stellar mass (see \citealt{Yang07} for more details, but note that they used the color-M/L ratio relation from Bell et al. 2003 to estimate stellar masses). We use the latter $M_{h}$ due to the group's stellar mass being a better constraint than luminosity (\citealt{More11}). \citet{Yang07} converted $M_{h}$ into $M_{200}$, defined as the mass enclosed within the group virial radius $R_{200}$ (at which the average group density is 200 times higher than the critical density). 

Finally, we cross-match the SDSS data with the galaxy morphology catalogue of \citet{Lintott11} by matching the SDSS J2000.0 position-based designation of each source. \citet{Lintott11} presented the data release of the Galaxy Zoo project\footnote{http://zoo1.galaxyzoo.org/}, which consists of an online tool that enables citizen scientists to visually classify SDSS galaxies. Through Galaxy Zoo each galaxy was visually classified by a median of 39 citizen scientists (with a minimum of 20). The raw results were de-biased (e.g. for the effect of higher-redshift galaxies appearing smoother as the morphological structure becomes blurred) and compared to a subset of expert classifiers. \citet{Bamford09} assigned each galaxy a probability of being an early-type galaxy (elliptical+S0) $P_{\rm{ell}}$, or a spiral/disc (clockwise, anticlockwise or edge-on spiral) galaxy, $P_{\rm{s}}$. We follow previous Galaxy Zoo studies (e.g. \citealt{Bamford09,Schawinski10,Masters10}) and apply a probability cut of 0.8 to identify elliptical and discs galaxies. 

By joining the \citet{Yang07} galaxy group and Galaxy Zoo catalogues we generate a sample of 127,780 galaxies in the redshift range $0.02<z<0.1$ and stellar mass range $10^{8}-10^{11.7}\Msun$. This sample contains both central and satellite galaxies, when selecting central galaxies only the stellar mass range changes to $10^{9}-10^{11.7}\Msun$. Fig.~\ref{SDSS_sample} shows the stellar masses of the sample as a function of redshift. The left panel shows the number of galaxies in the stellar mass-redshift plane, whereas the right panel shows the distribution of the sample in stellar mass bins. This sample not only has a halo mass assigned to each individual galaxy (as well as a central/satellite identification), but also a morphological classification. We find that from the sample of 127,780 galaxies, only 48,245 galaxies have a probability of being a disc or elliptical larger than $80\%$, meaning that roughly $60\%$ of galaxies do not show a clear morphology, and are thus classified as irregulars.

\citet{Yang07} estimated the halo masses of galaxy groups down to a minimum of $10^{11.6}\Msun$. Those galaxies that are missing halo mass estimates and/or morphology determinations are discarded. Throughout this work, however, we focus on central galaxies, which we define as the most massive galaxies from each group. Therefore the original sample of 127,780 galaxies is reduced to a sample of 93,160 central galaxies in the redshift range $0.02<z<0.1$ and stellar mass range $10^{9}-10^{11.7}\Msun$. When we apply the probability cut of 0.8 for galaxies to be either discs or ellipticals, the subsample of 93,160 central galaxies is further reduced to 36,736 galaxies. 

\subsection{Completeness}\label{Completeness}

The SDSS galaxy sample is more than $99\%$ complete in the stellar mass range $10^{9}-10^{12}\Msun$ and redshift range $0.02<z<0.1$ (\citealt{Strauss02}). However, our subsample of central galaxies does not have this same completeness due to missing halo/morphology determinations. To estimate the completeness of our sample, we therefore calculate the ratio between the galaxy stellar mass function (hereafter GSMF) calculated with our subsample and the GSMF estimates from \citet{Peng10}, \citet{Baldry12} and \citet{Weigel16}, and determine the stellar mass range where our GSMF exceeds 0.75 times the \citet{Peng10}, \citet{Baldry12} and \citet{Weigel16} GSMFs. We find that the completeness of our sample of central discs is larger than $75\%$ in the stellar mass range $10^{9.8}-10^{11}\Msun$, whereas central ellipticals are more than $75\%$ complete in the mass range $10^{9.8}-10^{11.6}\Msun$. For both discs and ellipticals, the incompleteness at low masses is due to missing halo mass estimates, while for discs the lack of a robust morphological classification produces a low completeness at high masses. We refer the reader to Appendix A for further details on the GSMF determinations, comparisons as well as completeness analysis.

In the following sections we investigate the SHMR and its dependence on galaxy morphology using the 36,736 SDSS central galaxies in the redshift range $0.02<z<0.1$ and stellar mass range $10^{9}-10^{11.7}\Msun$. We remind the reader, however, that the range of $>75\%$ completeness lies in the stellar mass range $10^{10}-10^{11}\Msun$ (which corresponds to halo masses of $\sim 10^{12}\Msun$).

\begin{figure} 
\includegraphics[angle=0,width=0.47\textwidth]{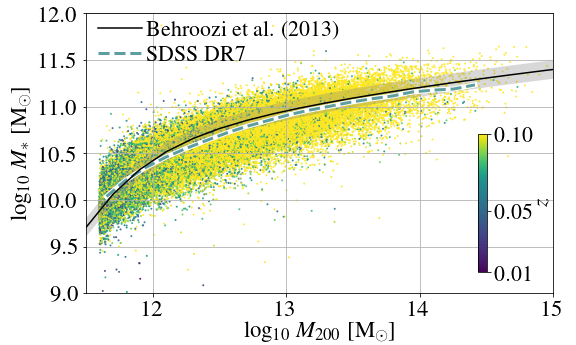}
\caption{Stellar-to-halo mass relation for 93,160 SDSS central galaxies. The color indicates the spectroscopic redshift of each galaxy (with lighter colors corresponding to higher redshift). The green dashed line shows the median relation, whereas the black solid line shows the best-fitting relation of Behroozi et al. (2013) obtained from abundance matching to observations, with the shaded region highlighting the 0.1 dex uncertainty.}
\label{SHMR}
\end{figure}

\subsection{Galaxy stellar-to-halo mass relation}\label{SDSS_SHMR}

Fig.~\ref{SHMR} shows the SHMR, with the green dashed line highlighting the median relation and the black solid line the best-fitting relation of \citet{Behroozi13} obtained from abundance matching to observations. Each dot in the figure corresponds to a galaxy coloured according to its spectroscopic redshift. The figure shows very good agreement between the median relation of our sample and that of \citet{Behroozi13}.

\begin{figure} 
\includegraphics[angle=0,width=0.47\textwidth]{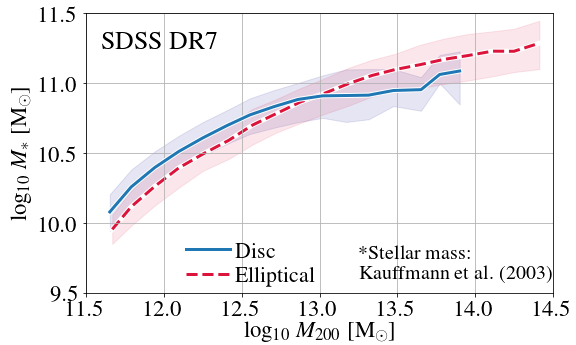}\\
\includegraphics[angle=0,width=0.47\textwidth]{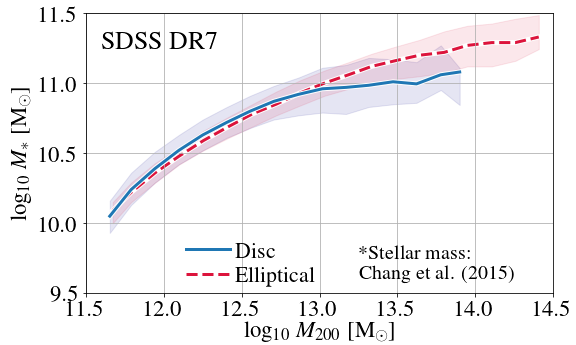}\\
\caption{SHMR for central disc (blue solid line) and elliptical galaxies (red dashed line), the blue and red shaded regions correspond to the $16-84$th percentiles. {\it Top}: The SHMR was calculated using the stellar mass estimates from Kauffmann et al. (2003). For $10^{11.7}$ to $10^{12.9}\Msun$ haloes, disc galaxies show a larger median stellar mass than ellipticals, with up to a factor of 1.4 difference for galaxies in $10^{12}\Msun$ haloes. {\it Bottom}: Same as top panel, but using the stellar mass estimates from \citet{Chang15}. For the same galaxy sample, the difference between the two panels shows that the dependence of the SHMR on morphology may either be of physical origin or an outcome of biased mass-to-light ratios affecting the stellar mass estimates.}
\label{SHMR_morphology}
\end{figure}

We next split the sample into discs and ellipticals. The top panel of Fig.~\ref{SHMR_morphology} shows the median SHMR for disc galaxies (blue solid line) and for elliptical galaxies (red dashed line), the $16-84$th percentiles are highlighted. It can be seen that for haloes in the mass range $10^{11.7}$ to $10^{12.9}\Msun$, disc galaxies have a larger median stellar mass than elliptical galaxies that reside in same-mass haloes, with the stellar mass difference peaking at a factor of 1.4 for galaxies in $10^{12}\Msun$ haloes. However, this morphology dependence disappears if we re-calculate the SHMR using the stellar masses from the \citet{Chang15} catalogue (bottom panel of Fig.~\ref{SHMR_morphology}). It can be seen that the median relations for discs and ellipticals residing in same-mass haloes are now in very good agreement. A similar result is obtained when switching to the stellar masses calculated by \citet{Brinchmann04}. In haloes more massive than $10^{13}\Msun$ both panels of Fig.~\ref{SHMR_morphology} show that the morphology-stellar mass relation changes and at fixed halo mass the median stellar mass of elliptical galaxies is larger than that of their disc-type counterparts, regardless the stellar mass estimate used. 

This lack of agreement between the SHMRs using the {\it{same galaxy catalogue}} but different stellar mass estimates indicates that the apparent morphology dependence of the low-mass SHMR may either have a physical origin or be the outcome of biased mass-to-light ratios. We discuss this in detail in the following section. 

\citet{Mandelbaum16} also used the stellar masses from \citet{Kauffmann03}, but combined these with halo masses estimated from galaxy-galaxy lensing. They separated galaxies according to their $g-r$ color, with galaxies with $g-r\ge 0.8$ classified as red and galaxies with $g-r< 0.8$ as blue, and found that at fixed stellar mass, red galaxies reside in haloes that are at least twice as massive as those haloes hosting blue galaxies. 

\begin{figure} 
\includegraphics[angle=0,width=0.47\textwidth]{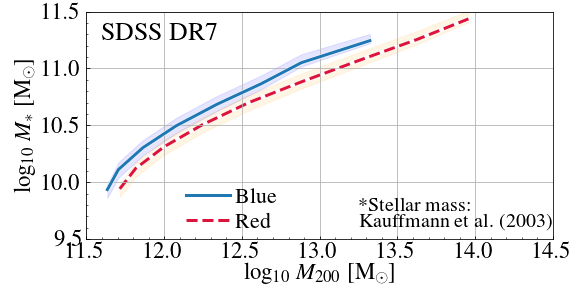}
\caption{SHMR for blue and red central galaxies shown as blue solid and red dashed lines, respectively. The light shaded regions show the $16-84$th percentile ranges. These relations are calculated using the stellar mass estimates from Kauffmann et al. (2003). Note that although the relations are plotted as stellar mass as a function of halo mass, the medians are calculated in bins of stellar mass in order to facilitate comparison to Mandelbaum et al. (2016). The figure shows that at fixed stellar mass, blue galaxies reside in lower mass haloes than their red counterparts, with the difference being larger than a factor of 2 in halo mass for galaxies with stellar masses $\geq 10^{11}\Msun$. This is in good agreement with the color-SHMR reported by \citet{Mandelbaum16}.}
\label{SHMR_color}
\end{figure}

We compare with the results of \citet{Mandelbaum16}, who calculated the color-SHMR using the $g-r$ color classification. We note, however, that \citet{Mandelbaum16} plotted the relation in stellar mass bins, rather than in halo mass bins as we have done for Fig.~\ref{SHMR_morphology}. Therefore, although we plot the stellar mass as a function of halo mass, we calculate the median SHMR in bins of stellar mass. This is shown in Fig.~\ref{SHMR_color}, where the median relations for blue and red galaxies are plotted as blue solid- and red dashed lines, respectively. It can be seen from the figure that at fixed stellar mass, blue galaxies reside in lower mass haloes than their red counterparts, with the difference being larger than a factor of 2 in halo mass for galaxies with stellar masses $\geq 10^{11}\Msun$. This is in very good agreement with \citet{Mandelbaum16}. If, on the contrary, the color-SHMR is calculated in halo mass bins, the relation changes. At fixed halo mass, blue galaxies have slightly larger stellar masses (by up to a factor of 1.2 in $10^{12}\Msun$ haloes) than their red counterparts. For haloes more massive than $10^{13}\Msun$, the relation changes and red galaxies are more massive than blue galaxies at fixed halo mass.

We warn the reader that the color-SHMR may be biased below a stellar mass of $10^{10.3}\Msun$ and halo mass of $10^{12.2}\Msun$. This is because the sample is only complete in halo mass down to $M_{200}=10^{11.7}\Msun$ as shown in Fig.~\ref{SHMR}.


\subsection{Discussion}\label{Discussion}

\subsubsection{Impact of morphology probability cut and central/satellite classification}

In this section we have used an SDSS sample of 36,736 central galaxies and showed that disc galaxies are up to a factor of 1.4 more massive than elliptical galaxies residing in same-mass haloes. This difference occurs in the halo mass range $10^{11.7}-10^{13}\Msun$ and when the stellar masses calculated by \citet{Kauffmann03} are used. When we re-calculate the SHMR using the stellar masses from \citet{Brinchmann04} or \citet{Chang15}, the morphology-SHMR dependency disappears in the halo mass range $10^{11.7}-10^{13}\Msun$. 

Galaxies are classified as centrals if they are the most massive member of the group (which in $\approx 90\%$ of groups it also corresponds to being the most luminous, \citealt{Yang07}). However, previous studies have shown that in $10\%$ of $10^{12.5}\Msun$ groups the most massive galaxy is not the central, and this fraction increases with group mass reaching $45\%$ for $10^{14}-10^{14.5}\Msun$ groups (see e.g. \citealt{Skibba11, Hoshino15, Lange18}). To determine if the assumption of the most massive galaxy being the central affects our results, we `contaminate' the central galaxy sample by assuming that satellite galaxies were misclassified as centrals. 

In the $10^{12}\Msun$ halo mass bin we replace $10\%$ of centrals by their most massive satellites that reside in the same halo, for higher-mass haloes we follow the fraction reported by \citet{Lange18}, which increases with halo mass reaching $45\%$ in the $10^{14}\Msun$ halo mass bin. The morphology-SHMR shown in Fig.~\ref{SHMR_morphology} is robust to the central/satellite galaxy classification for $10^{11.7}$ to $10^{12.8}\Msun$ haloes, using either \citet{Kauffmann03} or \citet{Chang15} stellar masses. In $>10^{13}\Msun$ haloes, the median stellar mass of discs galaxies decreases by up to 0.2 dex with respect to the original relation. This is because for these halo masses, the central galaxy tends to be significantly more massive than its disc satellites. For elliptical satellites, the mass difference is smaller and the SHMR remains nearly unchanged. We conclude from this analysis that in ${>}10^{13}\Msun$ haloes, the relatively large fraction of possible central/satellite misclassifications may have significantly affected the morphology-SHMR. In fact, the change of sign of the difference between the SHMRs of ellipticals and discs above halo masses of $10^{13}\Msun$ may be partially caused by misclassifications. We therefore focus on lower mass haloes for which the results are robust. This test is shown and further discussed in Appendix~\ref{SHMR_dependence_cen_sat}. We also analyse the impact of central/satellite misclassifications on the color-SHMR shown in Fig.~\ref{SHMR_color}, and find that this relation does not change when the sample is contaminated by satellites. 

Another factor that may bias the results presented in the previous subsections, is the morphology classification. We have followed previous Galaxy Zoo studies and applied a probability cut of $80\%$ for a central galaxy being either an elliptical or a disc. We analyse how this probability cut impacts our results by decreasing the threshold from $80\%$ to $60\%$ and $40\%$, thus allowing more uncertain classifications to enter our sample. For decreasing probability cuts, the difference between the median stellar masses of discs and elliptical galaxies slightly decreases. We find that for a probability cut of $40\%$ ($60\%$), disc galaxies show a larger median stellar mass than ellipticals, with up to a factor of 1.25 (1.3) difference for galaxies in $10^{12}\Msun$ haloes. Differently, the morphology-SHMR in $>10^{13}\Msun$ haloes changes by a larger factor. From this analysis we conclude that the morphology-SHMR in $<10^{13}\Msun$ haloes is robust to changes in the morphology probability cut. The changes of the SHMR with probability cut are shown in Appendix~\ref{SHMR_dependence_prob_cut}.


\subsubsection{Possible bias in mass-lo-light ratios}

The dependence of the SHMR with the stellar masses calculated by either \citet{Kauffmann03} or \citet{Chang15}, could be an indication of a possible bias in one or more of the derived mass-to-light ratios. It has generally been argued that stellar masses estimated for quiescent systems are more reliable than for star-forming ones (e.g. \citealt{Gallazzi09}). This is due to young stars outshining older stars, therefore hiding the old stellar populations and causing the color-M/L ratio relations to be uncertain for star-forming galaxies. In addition, star-forming galaxies contain more dust, which also contributes to the uncertainty in M/L ratios.

Derived M/L ratios depend on the assumed distribution of SFHs of the models used to interpret galaxies' SEDs. If simple SFHs (or single age models) are assumed, the estimated M/L ratios tend to be lower than the true ratios (e.g. \citealt{Pforr12}). The addition of bursts of star formation on top of a continuous SFH can produce M/L estimates systematically different by as much as $10\%$ to a factor of 2, depending on strength and fraction of the starbursts (e.g. \citealt{Bell01,Drory04,Pozzetti07,Gallazzi09,Wuyts09}).

\citet{Kauffmann03} modelled the $H\delta_{A}$ and $D_{n}4000$ spectral features measured from SDSS spectra in order to further constrain SFHs and M/L ratios. They showed that their M/L ratios strongly correlate with light concentration ($C$, defined as the ratio of the radii enclosing $90\%$ and $50\%$ of the petrosian $r$-band luminosity), a parameter that is higher ($C>2.6$) for elliptical galaxies and lower ($C<2.6$) for disc galaxies (\citealt{Strateva01}). More concentrated (elliptical) galaxies exhibit higher mass-to-light ratios than less concentrated (disc) galaxies. 

Differently, \citet{Brinchmann04} and \citet{Chang15} constrained the SFHs directly from fits to the SDSS galaxy spectra. The good agreement between the stellar masses from these studies seems to indicate that the addition of near-IR data does not necessarily yield more accurate stellar masses (\citealt{Taylor11}).

To further understand the morphology-SHMR, we resort to the EAGLE cosmological simulation in the following section.

\section{Eagle simulation}

The EAGLE cosmological hydrodynamical simulation (\citealt{Schaye15,Crain15}) has proven to broadly reproduce many properties of the observed galaxy population, such as galaxies' stellar masses (\citealt{Furlong15}), sizes (\citealt{Furlong17}), star formation rates and colours (\citealt{Trayford15,Trayford17}), and black hole masses and active galactic nuclei (AGN) luminosities (\citealt{Rosas16,McAlpine17}). \citet{Correa17} showed that EAGLE produces a galaxy population for which morphology is tightly correlated with the location in the colour-mass diagram, with red galaxies being mostly ellipticals and blue galaxies discs (see also \citealt{Trayford16,Correa19}). \citet{Matthee17} found that the scatter in the SHMR from EAGLE's central galaxies correlates strongly with halo concentration (or halo formation time), so that at fixed halo mass, a larger stellar mass corresponds to a more concentrated (and earlier forming) halo (see \citealt{Martizzi20} for a similar result from the IllustrisTNG simulations).

\subsection{Data}

The EAGLE reference model (Ref-L100N1504) is a cosmological, hydrodynamical simulation of 100 comoving Mpc on a side that was run with a  modified version of GADGET 3 (\citealt{Springel05}), a $N$-Body Tree-PM smoothed particle hydrodynamics (SPH) code with subgrid prescriptions for radiative cooling, star formation, stellar evolution, stellar feedback, black holes, and AGN feedback (see \citealt{Schaye15} for a detailed description). The Ref model contains 1504$^{3}$ dark matter (as well as gas) particles, with initial gas and dark matter particle masses of $m_{\rm{g}}=1.8\times 10^{6}\Msun$, $m_{\rm{dm}}=9.7\times 10^{6}\Msun$, respectively, and a Plummer equivalent gravitational softening of $\epsilon_{\rm{prop}}=0.7$ proper kpc at $z=0$. It assumes a $\Lambda$CDM cosmology with the parameters derived from {\it{Planck-1}} data (\citealt{Planck}), $\Omega_{\rm{m}}=1-\Omega_{\Lambda}=0.307$, $\Omega_{\rm{b}}=0.04825$, $h=0.6777$, $\sigma_{8}=0.8288$, $n_{s}=0.9611$, and primordial mass fractions of hydrogen and helium of $X=0.752$ and $Y=0.248$, respectively.

Dark matter haloes (and the self-bound substructures within them associated to galaxies) are identified using the Friends-of-Friends (FoF) and SUBFIND algorithms (\citealt{Springel01,Dolag}). Halo masses ($M_{200}$) are defined as all matter within the radius $R_{200}$ for which the mean internal density is 200 times the critical density. In each FoF halo, the `central' galaxy is the galaxy closest to the center (minimum of the potential), which is nearly always the most massive. The remaining galaxies within the FoF halo are its satellites. Following \citet{Schaye15}, we determine the galaxy stellar masses within spherical apertures of 30 proper kpc.

We calculate halo concentrations ($c_{200,\rm{DM}}$) from a dark matter only simulation that started from identical Gaussian density fluctuations as the Ref-L100N1504 model. We then identify the `same' haloes (that originate from the
same spatial locations) by matching the particles IDs in the two simulations, and fit NFW profiles (\citealt{Navarro97}) to the dark matter only spherically averaged density profiles. We measure the scale radius $r_{\rm{s}}$, that indicates where the logarithmic slope of the profile has the isothermal value of $-2$. Halo concentration is defined as the ratio between the virial radius and the scale radius, as $c_{200,\rm{DM}}\equiv R_{200}/r_{\rm{s}}$. It has been shown that $c_{200}$ strongly correlates with formation time, so that haloes that assemble earlier are more concentrated (e.g. \citealt{Wechsler02}).

We link dark matter haloes through consecutive snapshots following the merger trees from the EAGLE public database (\citealt{McAlpine16}). These merger trees were created using the D-Trees algorithm of \citet{Jiang14}, see also \citet{Qu17}. Using the merger trees we determine the halo formation time, $z_{\rm{f,halo}}$, defined as the redshift at which the halo mass reaches half of its $z=0$ mass. We also follow the galaxy assembly histories through 145 output redshifts between $z=0$ and $z=4$. This high time resolution is achieved by using the 145 RefL100N1504 `snipshots', which contain only the main particle properties but are output with much higher frequency than the regular snapshots. 

Finally, to quantify galaxy morphology, we follow \citet{Correa17} and use the fraction of stellar kinetic energy invested in ordered {\it{co-rotation}}, $\kappa_{\rm{co}}$. \citet{Correa17} showed that high-$\kappa_{\rm{co}}$ galaxies ($\kappa_{\rm{co}}\ge 0.4$) tend to be disc-shaped galaxies, whereas low-$\kappa_{\rm{co}}$ galaxies ($\kappa_{\rm{co}}< 0.4$) tend to be more spherical. After an extensive visual inspection of the Ref-L100N1504 galaxy sample, they used $\kappa_{\rm{co}}=0.4$ to separate galaxies that look disky from those that look elliptical. \citet{Thob17} showed that $\kappa_{\rm{co}}$ is tightly correlated with the major-to-minor axis ratio for EAGLE galaxies. Other works have shown that $\kappa_{\rm{co}}$ strongly correlates with various morphology metrics, such as angular momentum, bulge-to-total (disc-to-total) fractions, circularity, Gini coefficient (e.g. \citealt{Snyder15,Correa19,Trayford19,Thob19,Bignone20}). Recently, \citet{Bignone20} has confirmed that the simple threshold at $k_{\rm{co}}$ is enough to separate the transition between optically bulge dominated and disc dominated galaxies. 

\begin{figure} 
\centering
\includegraphics[angle=0,width=0.5\textwidth]{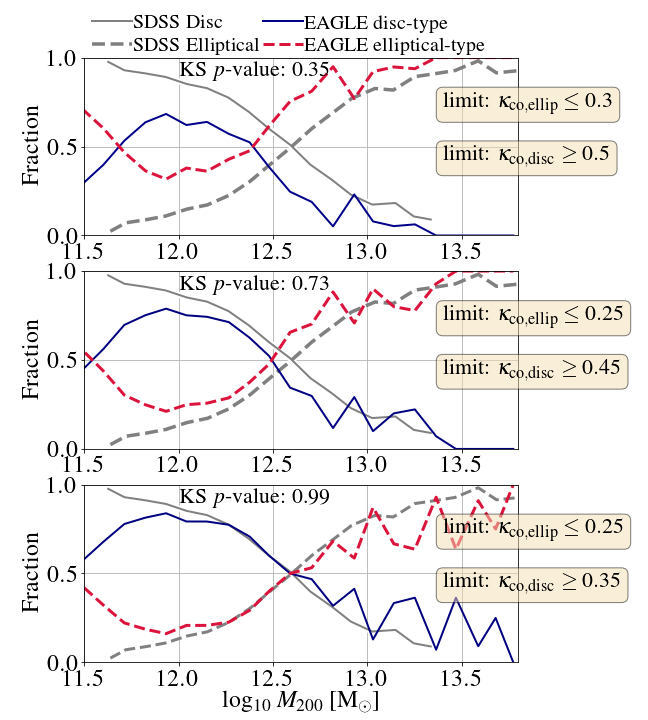}
\caption{Fraction of disc and elliptical central galaxies as a function of halo mass. Solid and dashed grey lines show the fractions for the SDSS galaxy sample described in Section 2.1. Solid blue and dashed red lines show the fractions for EAGLE galaxies separated into disc- and elliptical-type using the critical co-rotation parameter values $\kappa_{\rm{co}}=0.3,$ $0.5$ (top panel), $\kappa_{\rm{co}}=0.25,$ $0.45$ (middle panel) and $\kappa_{\rm{co}}=0.25,$ $0.35$ (bottom panel). The panels show that as the halo mass increases the fraction of elliptical galaxies increases while the fraction of discs galaxies decreases. This is however not the case for EAGLE galaxies in the halo mass range $10^{11.5}-10^{12}\Msun$, for which the galaxies are less well resolved. A Kolmogorov-Smirnov test indicates that the two galaxy samples (EAGLE and SDSS) residing in haloes more massive than $10^{12}\Msun$ are very similar ($p$-value 0.99) for the morphological cut of $\kappa_{\rm{co}}=0.25,$ $0.35$.}
\label{Comparison_SDSS_EAGLE}
\end{figure}

\subsection{Kinematic morphological indicator}

A stellar kinematic indicator provides a physically motivated morphological classification (e.g. \citealt{Fall83,Kormendy93,Kormendy04,Snyder15,Teklu15}). Although it may occasionally fail to discriminate between objects with different photometric morphologies, it correlates even more strongly with colour (\citealt{Emsellem07,Emsellem11,Thob19}). In this section we investigate whether a fixed $\kappa_{\rm{co}}$ cut produces a galaxy distribution of discs and ellipticals similar to that of the SDSS sample. To do so, we compare the fraction of disc and elliptical central galaxies in bins of halo mass. Fig.~\ref{Comparison_SDSS_EAGLE} shows the fraction of disc (solid grey lines) and elliptical (dashed grey lines) SDSS galaxies, as well as the fraction of disc- (solid dark blue lines) and elliptical-type (dashed red lines) EAGLE galaxies, that are separated into discs/ellipticals according to the kinematic indicator $\kappa_{\rm{co}}$, whose critical value we vary from $\kappa_{\rm{co,ellip}}\leq 0.3$ for ellipticals and $\kappa_{\rm{co,disc}}\geq 0.5$ for discs (top panel), to $\kappa_{\rm{co,ellip}}\leq 0.25$ and $\kappa_{\rm{co,disc}}\geq 0.45$ (middle panel), and to $\kappa_{\rm{co,ellip}}\leq 0.25$ and $\kappa_{\rm{co,disc}}\geq 0.35$ (bottom panel). Galaxies between the $\kappa_{\rm{co}}$ thresholds are considered `unclear' galaxies and not included in the analysis.

The panels show that as the halo mass increases, the fraction of elliptical SDSS galaxies increases from 0.1 in $10^{12}\Msun$ haloes to 0.95 in $10^{13.5}\Msun$ haloes. The opposite behaviour occurs for the fraction of disc SDSS galaxies, and both fractions reach 0.5 in $10^{12.6}\Msun$ haloes. EAGLE galaxies follow a similar behaviour as SDSS galaxies in the halo mass range $10^{12}-10^{13.5}\Msun$, but at lower halo masses the fraction of disc galaxies decreases while the fraction of elliptical galaxies increases. This is likely due to resolution effects, $10^{11.5}\Msun$ haloes host galaxies less massive than $10^{10}\Msun$ that therefore contain less than $10^{4}$ star particles. \citet{Schaye15} showed that resolution effects cause an upturn in the passive fraction at lower masses.

We vary $\kappa_{\rm{co}}$ to investigate which value yields a distribution of galaxies that is most similar to the observational sample, which used a photometric morphology classification. We perform a Kolmogorov-Smirnov (KS) test on the two galaxy populations: EAGLE (that depends on the $\kappa_{\rm{co}}$ cut) and SDSS. We find that for $\ge 10^{12}\Msun$ haloes, the distribution of central galaxies separated by $\kappa_{\rm{co,ellip}}\leq 0.25$ and $\kappa_{\rm{co,disc}}\geq 0.35$ results in a KS $p$-value of 0.99, indicating that the differences in the distributions are not statistically significant. The KS $p$-value drops to less than 0.4 when EAGLE galaxies are morphologically classified according to $\kappa_{\rm{co,ellip}}\leq 0.3$ and $\kappa_{\rm{co,disc}}\geq 0.5$. We conclude that the thresholds $\kappa_{\rm{co,disc}}\geq 0.35$ for discs and $\kappa_{\rm{co,ellip}}\leq 0.25$ for ellipticals produces a similar distribution of disc- and elliptical-type EAGLE galaxies to that of SDSS, and will adopt these as the critical values.

\begin{figure} 
\includegraphics[angle=0,width=0.47\textwidth]{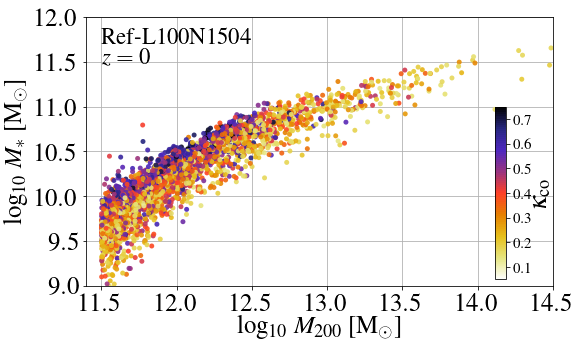}\\
\includegraphics[angle=0,width=0.47\textwidth]{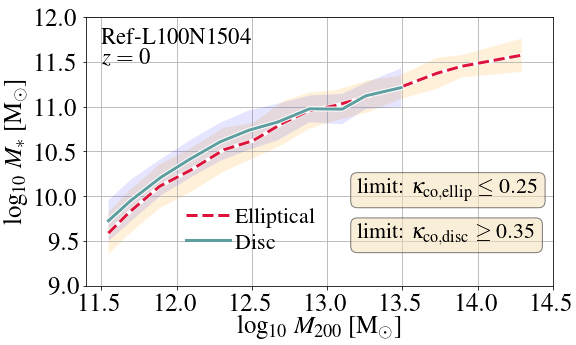}
\caption{{\it Top}: Relation between the stellar mass of $z=0$ central EAGLE galaxies and halo mass. Galaxies are coloured by $\kappa_{\rm{co}}$, a kinematic indicator of morphology. {\it{Bottom}}: Median SHMR for disc (solid blue line) and elliptical (red dashed line) central EAGLE galaxies. Galaxies are separated according to $\kappa_{\rm{co}}$ into discs ($\kappa_{\rm{co}}\ge 0.35$) and ellipticals ($\kappa_{\rm{co}}\leq 0.25$). The light blue and orange regions show the 16-84th percentile limits of the relation. In the halo mass range $10^{11.5}-10^{13}\Msun$, at fixed halo mass, disc galaxies are more massive than ellipticals.}
\label{SHMR_EAGLE}
\end{figure}

\subsection{Galaxy stellar-to-halo mass relation}

In this section we analyse the morphology-SHMR for EAGLE galaxies. The top panel of Fig.~\ref{SHMR_EAGLE} shows the relation between the stellar mass of $z=0$ central galaxies and halo mass. Galaxies are coloured by $\kappa_{\rm{co}}$. It can be seen that in the halo mass range $10^{11.5}-10^{13}\Msun$, at fixed halo mass disc galaxies tend to be more massive than their elliptical counterparts. The median relations in the bottom panel of the figure show that at fixed halo mass disc galaxies are up to a factor of 1.5 more massive than ellipticals. We obtain a similar result when galaxies are separated by fixed values of $\kappa_{\rm{co,disc}}=\kappa_{\rm{co,ellip}}=0.4$ or $\kappa_{\rm{co,disc}}=\kappa_{\rm{co,ellip}}=0.3$. This agrees well with the morphology-SHMR found in the SDSS galaxy sample with \citet{Kauffmann03} stellar masses shown in Fig.~\ref{SHMR_morphology}.

Recently, \citet{Moster19} analysed the SHMR for early- (passive) and late-type (active) galaxies defined according to a specific star formation rate threshold. They used an empirical model that followed the assembly history of haloes from a dark matter only simulation and was adjusted to reproduce GSMFs, star formation rates and non-star forming galaxy fractions. Differently from this work, they found that at fixed halo mass early-type (equivalent to ellipticals) are more massive than late-type galaxies (equivalent to discs). They reported these median trends when binning in halo mass, but noted than when the median trends where calculated in bins of stellar mass the relations changed, with late-type galaxies being more massive than early-types at fixed halo mass and over the halo mass range $10^{11}-10^{14}\Msun$. \citet{Moster19} argued that this was due to the scatter in the SHMR. 

We calculated the median SHMR relations for disc and elliptical galaxies by binning in stellar mass rather than halo mass, but this did not affect our conclusion that at fixed halo mass (with $M_{200}<10^{13}\Msun$) disc galaxies are more massive than elliptical galaxies. In Section~\ref{Physical_origin} we investigate whether halo assembly history or feedback from the central black hole can explain the morphology-SHMR of EAGLE galaxies.

\subsection{Comparison between EAGLE and SDSS}

\begin{figure} 
\includegraphics[angle=0,width=0.45\textwidth]{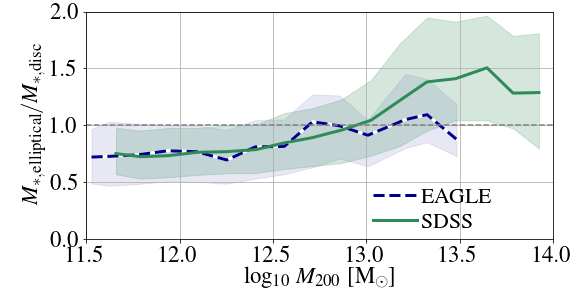}
\caption{Ratio between the median masses of elliptical and disc central galaxies, expressed as $M_{*,\rm{elliptical}}/M_{*,\rm{disc}}$, as a function of halo mass. The median ratios for EAGLE galaxies are shown using a dashed blue line and for SDSS galaxies using a green solid line, the shaded regions show the 16-84th percentile limits of the relations. The stellar masses of Kauffmann et al. (2003) are used for the SDSS sample. The figure shows very good agreement between the median ratios of elliptical and disc central galaxies from EAGLE and SDSS, but only for galaxies residing in haloes less massive than $10^{13}\Msun$. In higher-mass haloes the stellar masses of EAGLE disc and elliptical galaxies agree, whereas SDSS elliptical galaxies are more massive than discs at fixed halo mass.}
\label{Comparison_EAGLE_SDSS_morpho}
\end{figure}

The morphology-SHMR for EAGLE galaxies in $<10^{13}\Msun$ haloes agrees very well with the morphology-SHMR found in the SDSS galaxy sample with \citet{Kauffmann03} stellar masses. This can be seen in Fig.~\ref{Comparison_EAGLE_SDSS_morpho}, which shows the ratio between the median masses of elliptical and disc central galaxies, expressed as $M_{*,\rm{elliptical}}/M_{*,\rm{disc}}$, as a function of halo mass. The median ratios for EAGLE galaxies are shown by a dashed blue line, and by a green solid line for SDSS galaxies, the shaded regions show the 16-84th percentile limits of the relation. In higher-mass haloes ($>10^{13}\Msun$) the stellar masses of EAGLE disc and elliptical galaxies agree, whereas SDSS elliptical galaxies are more massive than discs at fixed halo mass.

\begin{figure} 
\includegraphics[angle=0,width=0.45\textwidth]{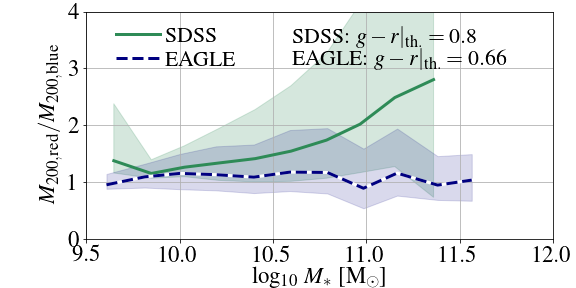}
\caption{Ratio between the median halo masses of red and blue central galaxies ($M_{200,\rm{red}}/M_{200,\rm{blue}}$) as a function of stellar mass. SDSS galaxies are separated into red and blue following the color-cut of $g-r=0.8$ of Mandelbaum et al. (2006) and the stellar masses of Kauffmann et al. (2003) are used. EAGLE galaxies are separated into red and blue following the color-cut of $g-r=0.66$. The median ratios for EAGLE galaxies are shown using a dashed blue line and for SDSS galaxies using a green solid line, the shaded regions show the 16-84th percentile limits of the relations. The figure shows that EAGLE does not reproduce the SDSS color-SHMR, EAGLE blue and red central galaxies reside in haloes of similar masses at fixed stellar mass. For SDSS, on the contrary, at fixed stellar mass, blue galaxies reside in lower-mass haloes than their red counterparts, with the difference being larger than a factor of 2 in halo mass for galaxies with stellar masses $\geq 10^{11}\Msun$.}
\label{Comparison_EAGLE_SDSS_color}
\end{figure}

Section~\ref{SDSS_SHMR} analyses the color-SHMR for SDSS galaxies, showing that at fixed stellar mass, blue galaxies reside in lower-mass haloes than their red counterparts, with the difference being larger than a factor of 2 in halo mass for galaxies with stellar masses $\geq 10^{11}\Msun$. This relation can be seen more clearly in Fig.~\ref{Comparison_EAGLE_SDSS_color}, which shows the ratio between the median halo masses of red and blue central galaxies ($M_{200,\rm{red}}/M_{200,\rm{blue}}$) from EAGLE and SDSS as a function of stellar mass. SDSS galaxies are separated into red and blue following the color-cut of $g-r=0.8$ and the stellar masses of \citet{Kauffmann03} are used (in both figures). Not many EAGLE galaxies have $g-r$ intrinsic colors larger 0.8 since the local minimum of the $g-r$ color PDF occurs at $g-r=0.66$, therefore we adopt this color-cut to separate galaxies into red and blue. It can be seen from the figure that EAGLE does not reproduce the SDSS color-SHMR, EAGLE blue and red central galaxies reside in haloes of similar masses at fixed stellar mass.

Due to the relative small scatter in the EAGLE SHMR it is possible to invert the relation, compare the ratio in stellar masses of red/blue galaxies as a function of halo mass, and find that the mean stellar masses of blue and red galaxies agree. This is an indication that while there is a correlation between galaxy morphology and color (as shown in \citealt{Correa17} for EAGLE galaxies), it does not necessarily hold for the SHMR. \cite{Matthee19} analysed the relation between star formation rate, stellar mass and halo mass using the EAGLE simulation. They showed that at fixed stellar mass, galaxies with relatively low star formation rates tend to reside in higher mass haloes. However, for stellar masses $M_{*}>10^{10}\Msun$ the correlation is rather weak and most of the scatter in the star formation-stellar mass relation is explained by black hole mass.

\section{Physical origin}\label{Physical_origin}

\subsection{Halo formation time}

The hierarchical assembly of dark matter haloes likely affects the morphology-SHMR. At fixed halo mass, galaxies residing in haloes that formed earlier tend to be more massive, not only because they have had more time for accretion and star formation (\citealt{Matthee17,Kulier19}), but also because the host haloes are more concentrated and thus have higher binding energies, making the galaxies' feedback less efficient (\citealt{Booth10,Davies19}). 

\begin{figure} 
\includegraphics[angle=0,width=0.47\textwidth]{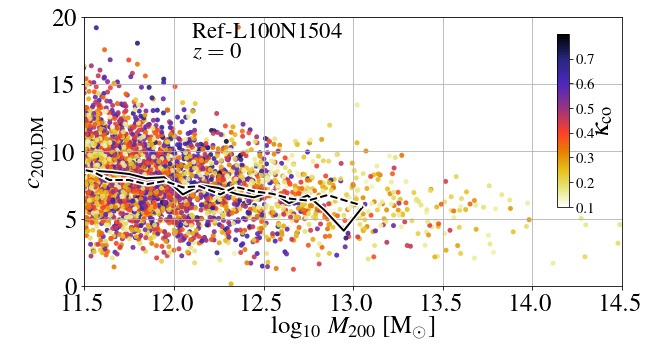}\\
\includegraphics[angle=0,width=0.47\textwidth]{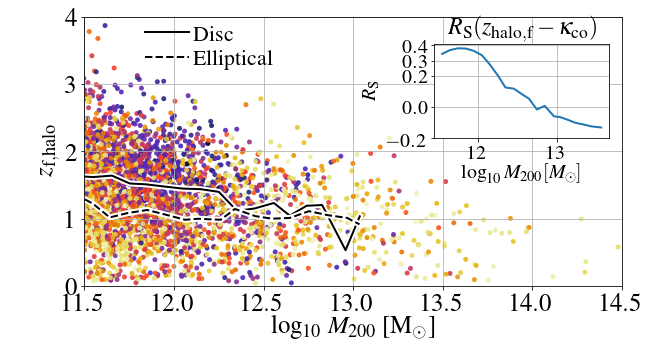}
\caption{{\it Top}: Relation between $z=0$ halo concentration, measured for matching haloes in the corresponding dark matter only simulation, halo mass and central galaxy morphology. Galaxies are coloured by their kinematic morphology, $\kappa_{\rm{co}}$. The thick solid and dashed lines indicate the median relations. {\it{Bottom}}: Halo formation time, $z_{\rm{f,halo}}$, as a function of halo mass. As in the top panel, galaxies are coloured by $\kappa_{\rm{co}}$ and the median relations are indicated. The inset shows the Spearman rank correlation coefficient ($R_{S}$) between $z_{\mathrm{f,halo}}$ and $\kappa_{\rm{co}}$ in bins of halo mass. In the halo mass range $10^{11.5}-10^{12}\Msun$ disc galaxies tend to reside in earlier forming haloes, but not in more concentrated haloes, than their elliptical counterparts.}
\label{zf_EAGLE}
\end{figure}

Fig.~\ref{zf_EAGLE} shows halo concentration (top panel) and halo formation time (bottom panel) as a function of halo mass. Dots in the figure correspond to $z=0$ central galaxies coloured by morphology, while the solid and dashed lines indicate the median relations. The inset of the bottom panel also shows the Spearman rank correlation coefficient ($R_{S}$) of the $z_{\rm{f,halo}}-\kappa_{\rm{co}}$ relation in bins of halo mass. Note that values larger (lower) than $R_{S}=	(-)0.3$ indicate that the (anti-)correlation is strong. 

From the bottom panel it can be seen that disc galaxies tend to reside in earlier forming haloes than their elliptical counterparts that reside in same-mass haloes. This is quantified in the inset, which shows a strong correlation between $z_{\rm{f,halo}}$ and $\kappa_{\rm{co}}$ in ${\sim}10^{12}\Msun$ haloes. Note that this also holds for smaller haloes (with masses between $10^{11.5}-10^{12}\Msun$). For ${\sim}10^{12}\Msun$ haloes, nevertheless, the median relations show that disc galaxies reside in haloes that formed around $z_{\rm{f,halo}}\approx 1.7$, whereas elliptical galaxies reside in haloes that formed 2 Gyr later, at around $z_{\rm{f,halo}}\approx 1.2$. To be able to reach the same mass as the haloes hosting discs, haloes hosting elliptical galaxies must have experienced a higher rate of mass growth, possibly explaining the $z=0$ morphological shape of their central galaxies.

For haloes more massive than $10^{12.1}\Msun$, $z_{\rm{f,halo}}$ does not seem to impact the morphology-SHMR as strongly. Interestingly, the correlation between galaxy morphology and halo formation time is not present in the halo concentration-mass plane. Disc galaxies reside in haloes with similar dark matter only concentrations as ellipticals.
 
\subsection{Central black hole}

An important aspect of the assembly history of a galaxy is the co-evolution between the galaxy and its supermassive black hole (hereafter BH). The energetic feedback released by the central BH not only self-regulates the growth of the BH itself, but also that of its host galaxy (e.g. \citealt{Silk98}). Observations report that the mass of the central BH correlates strongly with bulge mass (e.g. \citealt{Tremaine02}), indicating that elliptical galaxies tend to host higher-mass BHs than same-mass disc galaxies (e.g. \citealt{Watabe09,Shankar19}). A more massive BH has over time provided more energy to transport baryons, reducing the halo gas fraction, gas accretion and star formation (e.g. \citealt{Davies19,Oppenheimer20}), therefore lowering the galaxy stellar mass (for a given halo mass) and possibly explaining the morphology-SHMR. Note that in this section, when we analyse the impact of the central BH on the morphology-SHMR, we do not investigate whether the BH determines the galaxies' morphology, but rather whether the BH feedback lowers the galaxy's stellar mass for a given halo mass and morphology.

\begin{figure} 
\includegraphics[angle=0,width=0.47\textwidth]{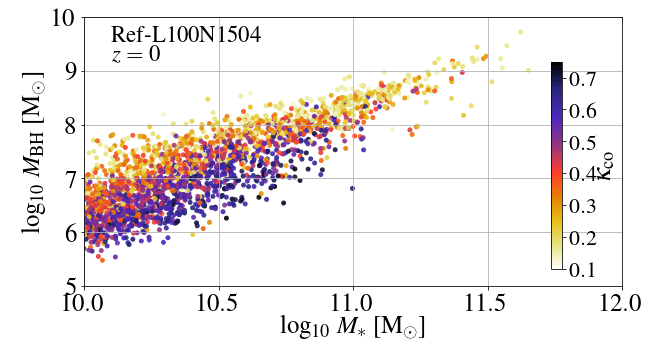}\\
\includegraphics[angle=0,width=0.47\textwidth]{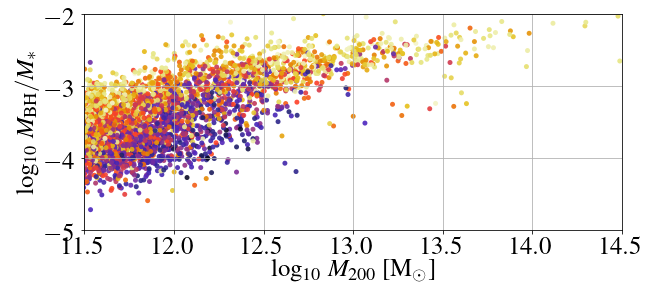}
\caption{{\it Top}: $z=0$ central BH mass-stellar mass relation, with galaxies coloured according to their kinematic morphology $\kappa_{\rm{co}}$. {\it{Bottom}}: ratio between the BH mass and the stellar mass as a function of halo mass. The panels show that central disc galaxies host less massive central BHs than their elliptical counterparts that are either of the same stellar mass (top panel) or residing in same-mass haloes (bottom panel).}
\label{BH_EAGLE}
\end{figure}

The top panel of Fig.~\ref{BH_EAGLE} shows the $z=0$ central BH mass-stellar mass relation, whereas the bottom panel shows the ratio between the BH and stellar masses as a function of halo mass. Galaxies are coloured according to $\kappa_{\rm{co}}$, showing that disc galaxies tend to host less massive BHs than elliptical galaxies of the same stellar mass (top panel). Similarly, at fixed halo mass, the ratio of BH mass and stellar mass is lower for disc galaxies than for ellipticals (bottom panel). This seems to indicate that the energetic outflows from the central BH prevented the further growth in mass of elliptical galaxies, possibly producing the morphology-SHMR.

To further investigate this we resort to a cosmological simulation of 50 comoving Mpc on a side where AGN feedback was switched off (hereafter named NoAGNL50N752 simulation). We compare the morphology-SHMR between galaxies from the NoAGNL50N752 and RefL50N752 simulations (Reference model run in a 50 Mpc box). 

\begin{figure} 
\includegraphics[angle=0,width=0.47\textwidth]{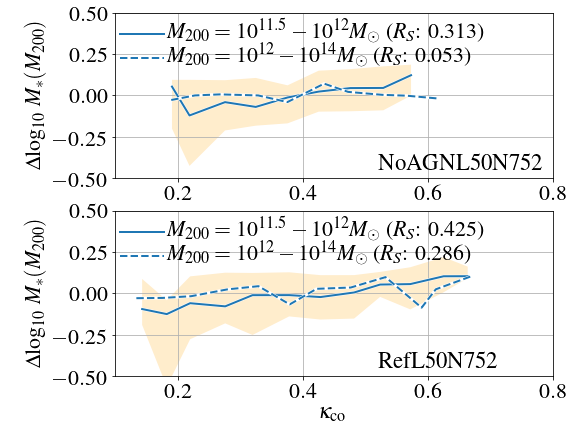}
\caption{{\it{Top \& Bottom}}: deviation from the median stellar mass given the halo mass ($\Delta\log_{10}M_{*}(M_{200})$) as a function of kinematic morphology $\kappa_{\rm{co}}$, for the NoAGNL50N752 and RefL50N752 simulations. In the panels the median relations of $\kappa_{\rm{co}}-\Delta\log_{10}M_{*}(M_{200})$ are calculated for two halo mass bins, $10^{11.5}-10^{12}\Msun$ (solid line) and $10^{12}-10^{14}\Msun$ (dashed line), with the Spearman rank correlation coefficient ($R_{S}$) of the relations indicated accordingly. The light orange region shows the 16-84th percentile limits of the low-mass bin relation. The panels show that for galaxies in haloes with masses between $10^{11.5}$ and $10^{12}\Msun$, the $\kappa_{\rm{co}}-\Delta\log_{10}M_{*}(M_{200})$ is strong for both simulations (with and without AGN feedback), indicating that for these halo masses the energetic feedback from the central BH does not produce the morphology-SHMR. For higher-mass haloes ($>10^{12}\Msun$), the $\kappa_{\rm{co}}-\Delta\log_{10}M_{*}(M_{200})$ is much weaker in the absence of AGN feedback, indicating that AGN feedback impacts on the morphology-SHMR.}
\label{BH_EAGLE_2}
\end{figure}

The top and bottom panels of Fig.~\ref{BH_EAGLE_2} show the deviation from the median stellar mass given the halo mass ($\Delta\log_{10}M_{*}(M_{200})$) as a function of $\kappa_{\rm{co}}$ for the NoAGNL50N752 (top) and RefL50N752 (bottom) simulations. The median relations of $\kappa_{\rm{co}}-\Delta\log_{10}M_{*}(M_{200})$ are calculated for two halo mass bins, $10^{11.5}-10^{12}\Msun$ (solid line) and $10^{12}-10^{14}\Msun$ (dashed line), and the Spearman rank correlation coefficients of the relations are indicated. The light orange region shows the 16-84th percentile limits of the $10^{11.5}-10^{12}\Msun$ bin relation.
 
The panels of Fig.~\ref{BH_EAGLE_2} show that there is a strong correlation between $\kappa_{\rm{co}}$ and $\Delta\log_{10}M_{*}(M_{200})$ (in both simulations) for galaxies residing in haloes with masses between $10^{11.5}-10^{12}\Msun$. This is quantified by the $R_{S}$ coefficients, with values of 0.313 and 0.425 for the NoAGNL50N752 and RefL50N752 models, respectively, which indicates that for these galaxies the energetic feedback from the central BH does not produce the morphology-SHMR. In higher-mass haloes ($>10^{12}\Msun$), there is no correlation between $\kappa_{\rm{co}}$ and $\Delta\log_{10}M_{*}(M_{200})$ in the NoAGNL50N752 model ($R_{S}=0.053$) and a weak correlation in the RefL50N752 model ($R_{S}=0.286$), from which we conclude that at these masses AGN feedback does impact the morphology-SHMR.

Interestingly, \citet{Bower18} and \citet{McAlpine18} showed that EAGLE BHs enter a rapid growth phase at a fixed critical halo virial temperature. However, if early-forming haloes (which we found tend to host disc galaxies) reached that critical temperature earlier than later-forming haloes (hosting ellipticals), why do disc galaxies host less massive BHs than ellipticals? The answer may be the rate of halo mass growth. Elliptical galaxies residing in later forming haloes likely experienced a faster rate of mass growth, that not only shaped the galaxies' morphologies into ellipticals, but also triggered a rapid growth phase of BHs (\citealt{McAlpine18}).

\begin{figure} 
\includegraphics[angle=0,width=0.47\textwidth]{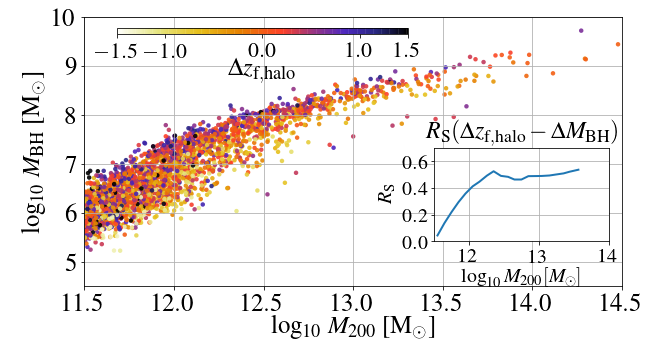}
\caption{$z=0$ central BH mass-halo mass relation, with galaxies coloured according to the deviation from the median halo formation time given the halo mass ($\Delta z_{\mathrm{f,halo}}$), so that $\Delta z_{\mathrm{f,halo}}>0$ ($\Delta z_{\mathrm{f,halo}}<0$) corresponds to earlier-forming (later-forming) haloes. The inset in the figure shows the Spearman rank correlation coefficient ($R_{S}$) between $\Delta z_{\mathrm{f,halo}}$ and $\Delta M_{\mathrm{BH}}$ (deviation from the median BH mass given the halo mass) in bins of halo mass. The figure shows that in the halo mass range $10^{11.5}-10^{12}\Msun$, there is no correlation between halo formation time and BH mass. Differently, for higher masses ($10^{12}-10^{14}\Msun$), there is a strong correlation between BH mass and halo formation time, indicating that earlier-forming haloes host more massive BHs than later-forming haloes.}
\label{BH_MH_zf}
\end{figure}

To better understand the role of BHs we analyse the $z=0$ central BH mass-halo mass relation shown in Fig.~\ref{BH_MH_zf}, where galaxies are coloured according to the deviation from the median halo formation time given the halo mass ($\Delta z_{\mathrm{f,halo}}$), so that $\Delta z_{\mathrm{f,halo}}>0$ ($\Delta z_{\mathrm{f,halo}}<0$) corresponds to earlier-forming (later-forming) haloes. The correlation between $\Delta z_{\mathrm{f,halo}}$ and $\Delta M_{\mathrm{BH}}$ (defined as the deviation from the median BH mass given the halo mass) in bins of halo mass is shown in the inset of the figure.

Fig.~\ref{BH_MH_zf} shows that in the halo mass range $10^{11.5}-10^{12}\Msun$ there is no correlation between halo formation time and BH mass. Differently, in the halo mass range $10^{12}-10^{14}\Msun$, there is a strong correlation between BH mass and halo formation time. This is quantified in the inset of the figure, which shows that the Spearman correlation coefficient $R_{S}$ is above 0.4. Fig.~\ref{BH_MH_zf} then shows that for haloes more massive than $10^{12}\Msun$, at fixed halo mass, larger BH masses correspond to earlier-forming haloes, and according to Fig.~\ref{BH_EAGLE}, more elliptical morphologies.

\subsection{Galaxy evolution}

In the previous sections we concluded that in the halo mass range $10^{11.5}-10^{12}\Msun$ it is the halo assembly history, and not the energetic feedback from the central BH, that produces the morphology-SHMR, but that at higher halo masses ($10^{12}-10^{12.5}\Msun$) AGN feedback impacts the galaxies' stellar masses, producing the morphology-SHMR.

\begin{figure*} 
\includegraphics[angle=0,width=\textwidth]{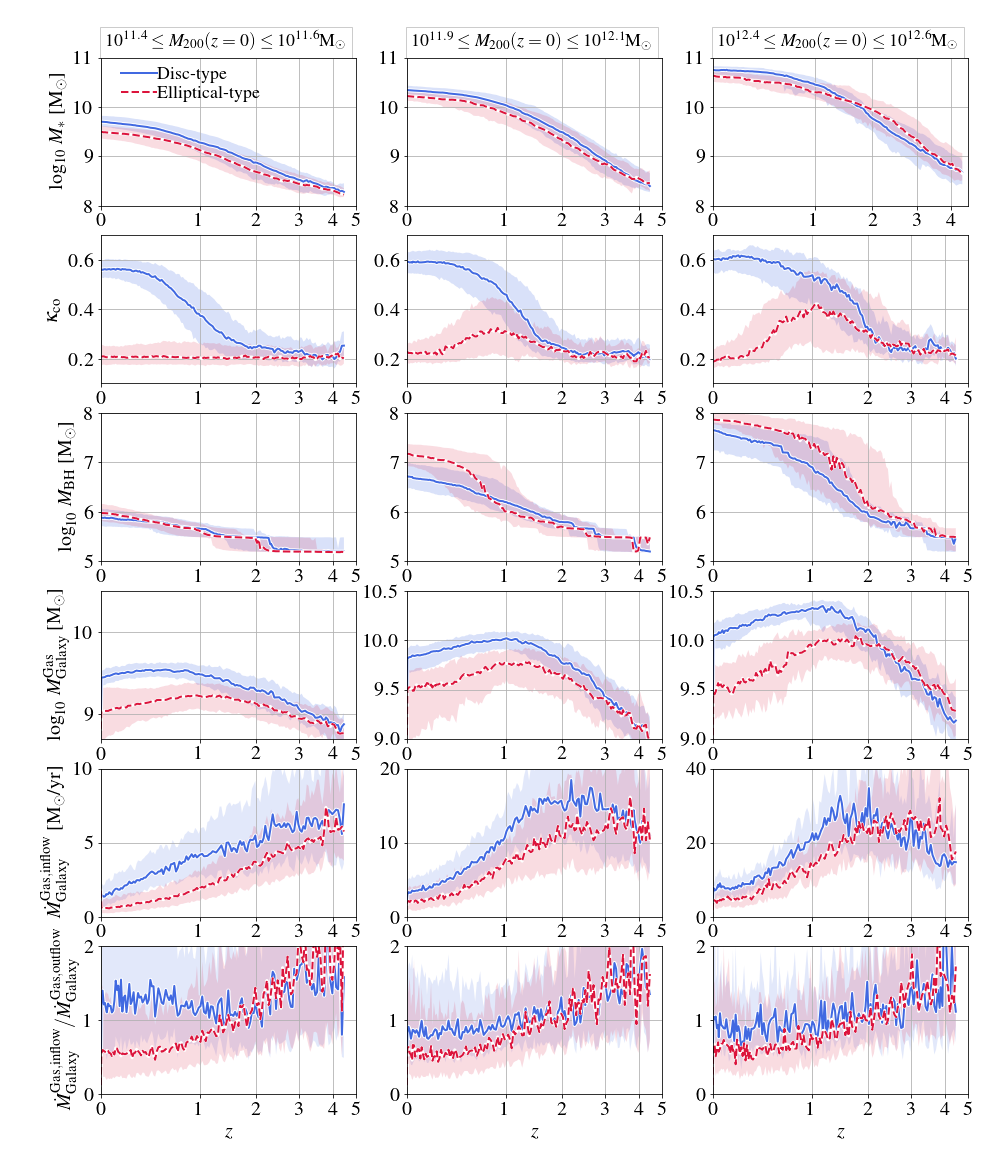}
\caption{{\it Top}: median stellar mass of discs (blue solid line) and ellipticals (red dashed line) from the low- (top-left), middle- (top-middle) and high- (top-right) halo mass sample  as a function of redshift. The following rows show, respectively, the kinematic morphology parameter, $\kappa_{\rm{co}}$, the median central BH mass, the total gas mass enclosed within $0.15\times R_{200}$, the median rate of gas accretion onto the galaxy, and the median ratio between the rates of galaxy gas inflow and outflow, as a function of redshift for the different halo mass samples.}
\label{GalaxyEvolution_EAGLE}
\end{figure*}

We investigate this further by following the mass assembly history of galaxies separated into three halo mass bins, referred to as the low-mass sample (galaxies in $10^{11.4}-10^{11.6}\Msun$ haloes), middle-mass sample ($10^{11.9}-10^{12.1}\Msun$ haloes) and high-mass sample ($10^{12.4}-10^{12.6}\Msun$ haloes). The top panels of Fig.~\ref{GalaxyEvolution_EAGLE} show the median stellar mass growth of discs (blue solid line) and ellipticals (red dashed line) from the low- (top-left), middle- (top-middle) and high-mass sample (top-right). In the low- and middle-mass samples, present-day disc galaxies were slightly more massive than present-day ellipticals throughout the redshift range 0-4, whereas in the high-mass sample, present-day elliptical galaxies were more massive until $z\approx 1.5$, when they were overtaken by the disc population. The second panels from the top show the morphological evolution of these samples. While elliptical galaxies from the low-mass sample were always ellipticals throughout the redshift range $0-4$, the middle and high-mass samples show that present-day elliptical galaxies developed a rotating disc at around $z\approx 1$, when the median values of $\kappa_{\rm{co}}$ reached values of 0.3 and 0.4, before turning into ellipticals. 

The third row from the top of Fig.~\ref{GalaxyEvolution_EAGLE} shows the central BH mass growth. In the low-mass sample there is no distinction between the BH masses of present-day discs and ellipticals, whereas in the middle-mass sample the BHs of present-day elliptical galaxies grew faster at $z<1$ than the disc-hosted BHs. The central BH of the present-day elliptical high-mass sample grew faster than the discs-hosted BH, even before the galaxies changed morphology. 

The bottom row, and the second and third rows from the bottom, show the ratio of gas inflows and outflows, the rate of gas inflow onto the galaxy and total gas mass in the galaxy as a function of redshift, respectively. For the low- and middle-mass galaxy samples these panels indicate that the rate of gas inflow has been larger for present-day disc galaxies than for present-day ellipticals over the redshift range 0-4. Disc galaxies have therefore had a larger gas fraction available for star formation than ellipticals. Interestingly, at $z>2$ both discs and ellipsoids have had a larger rate of gas inflow than outflow, but this changes for elliptical galaxies at $z<2$, where feedback has been more effective at generating outflows. 

We conclude that for central galaxies residing in haloes with masses between $10^{11.5}-10^{12}\Msun$, present-day disc galaxies are more massive than present-day ellipticals because they reside in earlier forming haloes, and hence have had not only more time for accretion and star formation, but also have had higher rates of gas inflow relative to outflows.

In galaxies residing in haloes with masses of $10^{12}-10^{12.5}\Msun$, BHs play a more dominant role in their evolution. For present-day elliptical galaxies, the faster growing black holes have ejected much of the halo's gas reservoir, reducing the rates of gas accretion onto galaxies as well as suppressing the (re)growth of a stellar disc.

\section{Summary}

We used SDSS DR7 data to construct a large sample of 127,780 galaxies (93,160 centrals, and 36,736 centrals with clear disc/elliptical morphologies) in the redshift range $0.02<z<0.1$  (Fig.~\ref{SDSS_sample}) that have a morphological classification (\citealt{Lintott08,Lintott11}), stellar mass measurements (\citealt{Kauffmann03,Brinchmann04,Chang15}), and halo mass estimates as well as central/satellite classifications (\citealt{Yang07}). We assessed the completeness of the sample, finding that the entire sample (as well as the centrals) is more than $75\%$ complete in the stellar mass range $10^{9}-10^{12}\Msun$ (Section~\ref{Completeness}).

We investigated the dependence of the galaxy stellar-to-halo mass relation (SHMR) on galaxy morphology for the SDSS sample and found that, in the halo mass range $10^{11.7}-10^{12.9}\Msun$, at fixed halo mass disc galaxies have a larger stellar mass than ellipticals, with up-to a factor of 1.4 difference for galaxies in $10^{12}\Msun$ haloes (Fig.~\ref{SHMR_morphology}). This was concluded when using the stellar mass estimates from \citet{Kauffmann03}, but when the stellar masses were changed to those calculated by \citet{Chang15} or \citet{Brinchmann04} the morphology-SHMR disappears for this halo mass range (Fig.~\ref{SHMR_morphology}). For halo masses larger than $10^{13}\Msun$, discs are less massive than ellipticals in same-mass haloes, regardless of whose stellar mass estimates we use. However, we found that in massive haloes the results for disc galaxies may be affected by central/satellite misclassifications. 

We have further investigated the SHMR by looking into the difference between the stellar and halo masses of galaxies separated by color. We calculated the relation in bins of stellar mass and found that at fixed stellar mass, blue galaxies reside in lower mass haloes than their red counterparts, with the difference being larger than a factor of 2 in halo mass for galaxies with stellar masses larger than $10^{11}\Msun$ (Fig.~\ref{SHMR_color}). 

We discussed the impact of the central/satellite classification in biasing our results, as well as the morphology probability cut. We have found that if a large fraction ($>10\%$) of central galaxies are satellites misclassified as centrals, the morphology-SHMR changes (in up to 0.2 dex) for haloes more massive than $10^{13}\Msun$. The color-SHMR, on the contrary, does not change. Similarly, changes in the cut of Galaxy Zoo assigned probabilities of galaxies being discs or ellipticals only affects the morphology-SHMR in $>10^{13}\Msun$ haloes. We also discussed the differences in the techniques used by \citet{Kauffmann03}, \citet{Brinchmann04} and \citet{Chang15} to measure galaxy stellar masses (Section~\ref{Discussion}). For higher halo masses ($>10^{13}\Msun$), discs have lower stellar masses than ellipticals in same-mass haloes, regardless of whose stellar mass estimate is used.

To understand the origin of the morphology-SHMR we turned to the EAGLE cosmological simulation and found the same morphology-SHMR as the one reported for the SDSS galaxies using the stellar masses of \citet{Kauffmann03}. EAGLE galaxies were separated according to $\kappa_{\rm{co}}$ (a stellar kinematics-based morphology indicator) into disc ($\kappa_{\rm{co,disc}}\ge 0.35$) and elliptical galaxies ($\kappa_{\rm{co,ellip}}<0.25$). We found that in the halo mass range $10^{11.5}-10^{13}\Msun$, at fixed halo mass, disc galaxies are more massive than their elliptical counterparts, with the median masses being up to a factor of 1.5 larger (Fig.~\ref{SHMR_EAGLE}). 

In the halo mass range $10^{11.5}-10^{12}\Msun$ EAGLE disc galaxies reside in earlier forming haloes than their ellipticals counterparts (Fig.~\ref{zf_EAGLE}). Disc galaxies may be more massive because they had more time for accretion and star formation, higher rates of gas inflow, as well as higher rates of inflow relative to outflows, than ellipticals (Fig.~\ref{GalaxyEvolution_EAGLE}). We also show that in this halo mass range, the energetic feedback from the central black hole (BH) is not responsible for the morphology-SHMR (Fig.~\ref{BH_EAGLE_2}), despite the fact that disc galaxies host less massive central BHs than their elliptical counterparts of the same stellar mass (Fig.~\ref{BH_EAGLE}).

We followed the assembly history of galaxies separated into different halo mass bins (Fig.~\ref{GalaxyEvolution_EAGLE}), from which we concluded that only for haloes between $10^{12}$ and $10^{12.5}\Msun$, elliptical galaxies are less massive than discs because of their central BHs, which grew faster, ejecting more of the gas reservoir, reducing star formation, and preventing the galaxy from growing in mass and (re-)growing a disc.

\section*{Acknowledgements}

We thank the anonymous referee for fruitful comments that improved the original manuscript. CC is supported by the Dutch Research Council (NWO Veni 192.020). CC acknowledges various public python packages that have greatly benefited this work: \verb|scipy| (Jones et al. 2001), \verb|numpy| (van der Walt et al. 2011), \verb|matplotlib| (Hunter 2007) and \verb|ipython| (Perez \& Granger 2007). This work used the DiRAC Data Centric system at Durham University, operated by the Institute for Computational Cosmology on behalf of the STFC DiRAC HPC Facility (www.dirac.ac.uk). This equipment was funded by BIS National E-infrastructure capital grant ST/K00042X/1, STFC capital grant ST/H008519/1, and STFC DiRAC Operations grant ST/K003267/1 and Durham University. DiRAC is part of the National E-Infrastructure.

\section*{Data availability}

The data supporting the plots within this article are available on reasonable request to the corresponding author.

\bibliography{biblio}
\bibliographystyle{mn2e}

\appendix

\section{Completeness analysis}

In this section we further describe the completeness analysis introduced in Section 2.2. The SDSS sample used in this work consists of 127,780 (93,160 centrals, and 36,736 with clear disc/elliptical morphology) galaxies in the redshift range $0.02<z<0.1$, which due to missing halo mass/morphology estimates is not entirely complete throughout the stellar mass range $10^{9}-10^{12}\Msun$. We define the completeness as the ratio between the GSMF derived from the sample and the GSMF estimates from \citet{Peng10}, \citet{Baldry12} and \citet{Weigel16}, and analyse the stellar mass range above which the completeness is larger than 0.75.  

We begin by constructing the GSMF following the $1/V_{\rm{max}}$ method (\citealt{Schmidt68}), where $V_{\rm{max}}$ is the maximum volume within which a galaxy at redshift $z_{i}$ with stellar mass $M_{*}$ can be detected. 

To calculate $V_{\rm{max}}$ for each galaxy, we first estimate the limiting mass, $M_{*,\rm{lim}}$, of each galaxy defined as the mass the galaxy would have if its apparent magnitude were equal to the limiting magnitude of the survey, which for SDSS is $m_{\rm{lim}}=17.77$ (\citealt{Strauss02,Abazajian09}). $M_{*,\rm{lim}}$ can be computed for each galaxy in the sample as $\log_{10}M_{*,\rm{lim}}=\log_{10}M_{*}+0.4\times(m-m_{\rm{lim}})$, where $M_{*}$ is the galaxy's stellar mass, and $m$ its apparent magnitude in the $r$-band. 

The stellar mass limit of the sample is derived by selecting the $M_{*,\rm{lim}}$ value of the $20\%$ faintest galaxies at each redshift, and defining $M_{*,\rm{min}}(z)$ as the upper envelope of the $M_{*,\rm{lim}}$ distribution below which $95\%$ of the $M_{*,\rm{lim}}$ values lie. The limit $M_{*,\rm{min}}(z)$ corresponds therefore to the $95\%$ completeness limit at each redshift observable by the survey.

\begin{figure*} 
\includegraphics[angle=0,width=0.85\textwidth]{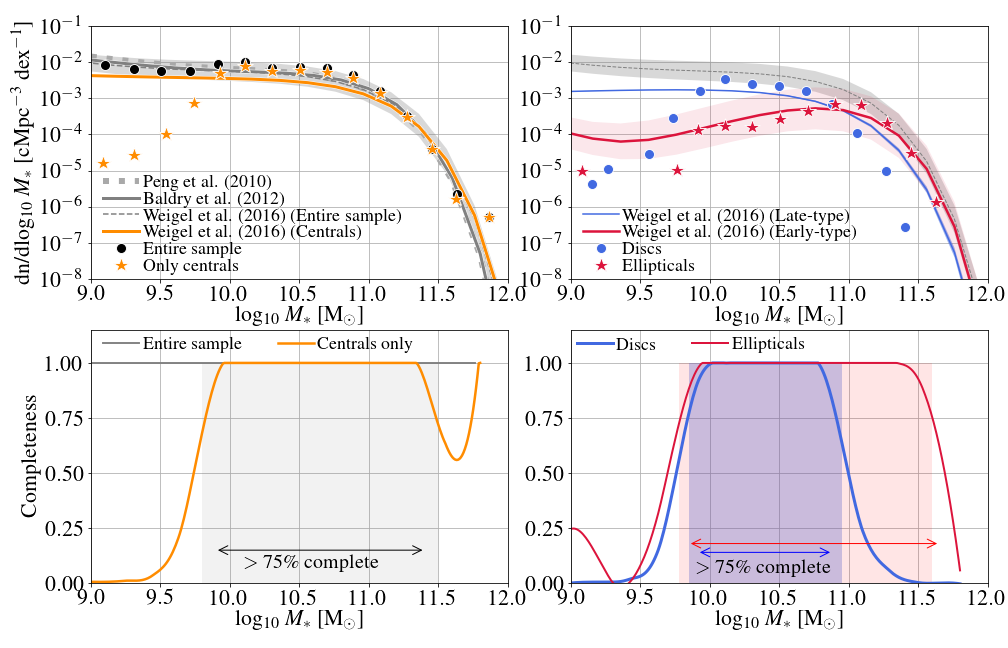}
\caption{{\it Top-left:} Comparison between the galaxy stellar mass functions (GSMF) derived from the SDSS galaxy sample that includes morphology and halo mass estimates, described in Section 2.1, and the GSMFs of Peng et al. (2010, who used the zCosmos survey shown in grey short-dashed line), Baldry et al. (2012, who used the GAMA survey shown in grey solid line) and Weigel et al. (2016, who used SDSS data shown in grey dashed line for all galaxies and solid orange line for centrals). The GSMFs constructed using the entire sample and only centrals are shown as black circles and orange star-symbols, respectively. {\it Top-right:} The GSMFs for disc and elliptical central galaxies are shown using blue circles and red star-symbols, respectively. The GSMFs of Weigel et al. (2016) are shown in blue and red solid lines for discs and ellipticals, respectively, with the red shaded region showing the $1\sigma$ uncertainty in the GSMF for ellipticals. {\it Bottom-left:} Completeness as a function of stellar mass for the entire sample (grey dashed line) and centrals (orange solid line). The grey shaded region shows the $>75\%$ mass range completeness of the central galaxies sample. {\it Bottom-right:} Completeness as a function of stellar mass for the subsamples of discs (blue solid line) and ellipticals (red thin-solid and dashed lines) central galaxies. The shaded regions highlight the stellar mass range where the samples are more than $75\%$ complete, i.e. the mass function exceeds 0.75 times the Weigel et al. (2016) GSMFs.}
\label{GSMF}
\end{figure*}

Next, we invert the $M_{*,\rm{min}}(z)$ relation to determine the maximum redshift, $z_{\rm{max}}$, out to which a galaxy with stellar mass $M_{*}$ can be detected. $z_{\rm{max}}$ is used in the $1/V_{\rm{max}}$ technique to weight each galaxy by the maximum detection volume. By doing so we are correcting for the Malmquist bias (\citealt{Malmquist22}), where faint, low-mass sources can only be detected in a small volume, while bright, massive sources can be detected in the entire sample volume. 

The GSMF, hereafter defined as $\Phi$, is calculated in 0.2 dex bins of stellar mass by summing all galaxies, $N_{\rm{gal}}$, in the mass bin as follows 

\begin{equation}
\Phi {\rm{d}}\log_{10}M=\sum_{i}^{N_{\rm{gal}}}\frac{w_{i}^{-1}}{V_{\rm{max},i}},
\end{equation}

\noindent where $w_{i}$ is the spectroscopic completeness of the source that we extract from the catalogue of \citet{Yang07} and $V_{\rm{max},i}$ is given by $V_{\rm{max},i}=\frac{4\pi}{3}\frac{\Omega_{s}}{\Omega_{\rm{sky}}}[d(z_{\rm{max},i})^{3}-d(z_{\rm{min},i})^{3}]$, with $\Omega_{\rm{sky}}=41,253$ deg$^{2}$ the surface area of the sky, $\Omega_{s}=7,748$ deg$^{2}$ the surface area covered by the sample, $d(z)$ the comoving distance to redshift $z$, $z_{\rm{min}}=0.02$ and $z_{\rm{max}}={\rm{min}}(0.1,z_{\rm{max},i})$. 

We compare the GSMF constructed from this sample with the GSMFs from \citet{Peng10}, \citet{Baldry12} and \citet{Weigel16}. \citet{Peng10} used the zCosmos spectroscopic survey (Lilly et al. 2007) to calculate GSMFs in the redshift range $0.1<z<2$. Their galaxy stellar masses were calculated assuming a \citet{Chabrier03} IMF. \citet{Baldry12} calculated the GSMF in the redshift range $0.002<z<0.06$ using an area of 143 deg$^{2}$ from the Galaxy And Mass Assembly (GAMA) survey DR1 (\citealt{Baldry10,Driver11}). Their GSMF was determined from a sample of 5,210 galaxies using a density-corrected maximum volume method, with stellar masses calculated in \citet{Taylor11} assuming a \citet{Chabrier03} IMF. Similar to this work, \citet{Weigel16} used a sample of $\sim$100,000 SDSS galaxies in the redshift range $0.02<z<0.06$. They applied the standard $V_{\rm{max}}$ method to calculate the GSMF using stellar masses from \citet{Kauffmann03} who assumed a \citet{Kroupa01} IMF. For consistency we convert their GSMF to a Chabrier IMF. \citet{Peng10}, as well as \citet{Baldry12} and \citet{Weigel16}, assumed a $\Lambda$CDM cosmology with $H_{0}=70$km s$^{-1}$ Mpc$^{-1}$ and $\Omega_{\rm{m}}=0.3$. We thus convert our $V_{\rm{max}}$ estimates to that cosmology for better comparison. The top-left panel in Fig.~\ref{GSMF} shows the GSMF constructed from this sample (with black circles corresponding to the entire sample and orange stars to central galaxies only), and the GSMFs from \citet[short-dashed line]{Peng10}, \citet[solid line]{Baldry12} and \citet[long-dashed line]{Weigel16}.

There is very good agreement between the GSMF for the entire sample and the GSMFs from the literature in the $10^{9}-10^{12}\Msun$ stellar mass range, this is highlighted in the bottom-left panel that shows the completeness of the entire sample (grey line), defined as the ratio between this work's GSMF and the GSMF of \citet{Baldry12}. Note that when we calculate the completeness we take into account the uncertainty in the best-fit parameters of the GSMFs from the literature, and plot the average range. 

We next estimate the completeness of central galaxies only, to do so the calculate the ratio between GSMF for centrals and the GSMFs from \citet{Weigel16} that was also calculated for central galaxies only. The bottom-left panel also shows the completeness for central galaxies in orange, and it can be seen that there is very good agreement between the GSMFs of our work and that of \citet{Weigel16} at the high-mass end, but at lower masses ($M_{*}<10^{9.8}\Msun$) the number of central galaxies largely decreases. This is due to galaxies residing in groups smaller than $10^{11}\Msun$ lacking halo mass estimates. We then conclude that we are more than $75\%$ complete for central galaxies in the stellar mass range $10^{9.8}-10^{11.5}\Msun$, as highlighted by the grey-shaded region.

An analysis of the completeness of central disc- and elliptical-type galaxies is shown in the right panels of Fig.~\ref{GSMF}. The top-right panel compares the GSMF for discs and ellipticals (shown in blue circles and red star symbols, respectively) with the GSMFs from \citet[shown in solid lines]{Weigel16}, who also used the morphological classification from Galaxy Zoo. The bottom-right panel shows the ratio between the two GSMFs for disc (blue line) and elliptical (red lines) galaxies. It can be seen that in the mass range $10^{9.9}-10^{10.9}\Msun$ the disc sample is more than $75\%$ complete. For the elliptical sample, the completeness is higher than $75\%$ in the $10^{9.8}-10^{11.6}\Msun$ stellar mass range. The panels show that we are missing massive disc galaxies, this is because these galaxies lack a morphological estimation of being discs with probability larger than $80\%$.

\section{Central/satellite classification}\label{SHMR_dependence_cen_sat}

In this section we investigate the impact of the central/satellite classification on our results by analysing how `contaminating' the sample with satellites, under the assumption that those were misclassified as centrals, changes the morphology-SHMR. We do so by first separating the sample into halo mass bins of 0.2 dex, for the $10^{12}\Msun$ halo mass bin we randomly replace $10\%$ of central galaxies by satellites that reside in the same halo and have the nearest stellar mass. For higher-mass haloes we follow the fraction reported by \citet{Lange18}, which increases with halo mass from $12\%$ in $10^{12.5}\Msun$ haloes to $45\%$ for $10^{14}\Msun$ haloes. We allow for up to 0.4 dex difference in stellar mass between the central and satellite, if that is not met we randomly select another central galaxy and look for its closest-in-stellar mass satellite. 

\begin{figure} 
\begin{center}
\includegraphics[angle=0,width=0.4\textwidth]{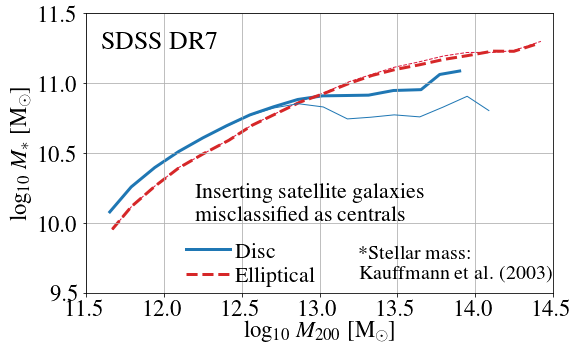}
\caption{SHMR for central disc (thin blue solid line) and elliptical galaxies (thin red dashed line) after in each halo mass bin of 0.2 dex, $10\%$ to $45\%$ of random central galaxies were replaced by their most massive satellites. The original SHMR (without central misclassification) is shown in thick blue solid and red dashed lines. The relations were calculated using the stellar mass estimates from Kauffmann et al. (2003). The figure shows that the morphology-SHMR relation is robust to the central/satellite galaxy classification for $10^{11.7}$ to $10^{12.8}\Msun$ haloes, for higher-mass haloes the relation still holds, with ellipticals being more massive than discs at fixed halo mass, but the median stellar masses of discs is lower ($<0.2$ dex) when satellite galaxies are misclassified as centrals.}
\label{cen_sat_impact}
\end{center}
\end{figure}

Fig.~\ref{cen_sat_impact} shows the SHMR for central disc (blue thin solid line) and elliptical galaxies (red thin dashed line) after the sample contamination. The original SHMR (presented in Section~\ref{SDSS_SHMR}) is shown in thick solid blue and dashed red lines for discs and ellipticals, respectively. We find that contaminating the sample by satellites changes the SHMR, but only for discs in haloes more massive than $10^{12.8}\Msun$. This is because in these haloes, the central galaxy tends to be significantly more massive than its disky satellites. When satellites replace centrals the median stellar mass for discs decreases by up to 0.2 dex. For $10^{11.7}$ to $10^{12.8}\Msun$ haloes the morphology-SHMR is robust to the central/satellite galaxy classification, using either \citet{Kauffmann03} or \citet{Chang15} stellar masses. We conclude from this analysis that in ${>}10^{13}\Msun$ haloes, the large fraction of possible central/satellite misclassifications affects the morphology-SHMR.

\section{Impact of morphology selection}\label{SHMR_dependence_prob_cut}
	
Section~\ref{Discussion} discusses how the probability cut used in the morphology classification biases the results presented in Section~\ref{SDSS_section}. We have followed previous Galaxy Zoo studies and applied a probability cut of $80\%$ for an individual central galaxy being either an elliptical or a disc. In this appendix we investigate whether our results are sensitive to this probability threshold, by decreasing the threshold from 80 to $60\%$ (this means increasing the central galaxy sample from 36,736 to 70,789 galaxies), and also to $40\%$ (increasing the central galaxy sample from 36,736 to 91,144 galaxies).

\begin{figure}
\begin{center}
\includegraphics[angle=0,width=0.4\textwidth]{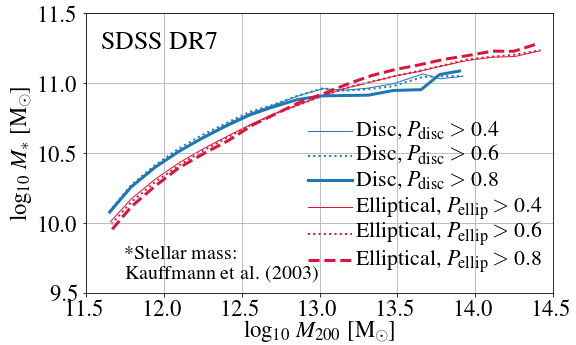}
\caption{SHMR for central disc (blue lines) and elliptical galaxies (red lines). The relations were calculated using the stellar mass estimates from Kauffmann et al. (2003). The panels show the morphology-SHMR after galaxies for which the assigned probability of being a disc/elliptical is greater than $40\%$, $60\%$ and $80\%$ are selected as such. The figure shows that the resulting morphology-SHMR is mostly affected by the probability cut in $>10^{13}\Msun$ haloes.}
\label{morpho_impact_1}
\end{center}
\end{figure}

Fig.~\ref{morpho_impact_1} shows the morphology-SHMR after galaxies with probabilities larger than $60\%$ and $40\%$ of being discs/ellipticals are selected as such. It can be seen that the resulting morphology-SHMR changes when different probability cuts are applied. For the probability cut of $80\%$, the median stellar masses of disc galaxies in $<10^{13}\Msun$ haloes are up to a factor of 1.4 higher than the median masses of their elliptical counterparts. When the probability cut is reduced to $60\%$ and $40\%$, the factor difference between the median masses of discs and ellipticals only decreases to 1.3 and 1.25, respectively. Differently, the morphology-SHMR in $>10^{13}\Msun$ haloes changes by a larger factor. From this analysis we conclude that the morphology-SHMR in $<10^{13}\Msun$ haloes is robust to changes in the morphology probability cut.

\end{document}